\documentclass[11pt,a4paper,english,amssymb,nofootinbib,superscriptaddress,showpacs]{revtex4}
\usepackage{lmodern}
\usepackage[T1]{fontenc}
\usepackage[latin9]{inputenc}
\usepackage{color}
\usepackage{babel}

\usepackage{amsmath}
\usepackage{amssymb}
\usepackage{esint}
\usepackage[unicode=true, pdfusetitle,
 bookmarks=true,bookmarksnumbered=false,bookmarksopen=false,
 breaklinks=false,pdfborder={0 0 1},backref=false,colorlinks=false]
 {hyperref}

\makeatletter
\@ifundefined{textcolor}{}
{%
 \definecolor{BLACK}{gray}{0}
 \definecolor{WHITE}{gray}{1}
 \definecolor{RED}{rgb}{1,0,0}
 \definecolor{GREEN}{rgb}{0,1,0}
 \definecolor{BLUE}{rgb}{0,0,1}
 \definecolor{CYAN}{cmyk}{1,0,0,0}
 \definecolor{MAGENTA}{cmyk}{0,1,0,0}
 \definecolor{YELLOW}{cmyk}{0,0,1,0}
 }

\@ifundefined{definecolor}
 {\usepackage{color}}{}
\usepackage{latexsym}\usepackage{bm}

\makeatother

\makeatother

\begin{document}

\title{AdS Waves as Exact Solutions to Quadratic Gravity}

\author{\.{I}brahim Güllü }

\email{e075555@metu.edu.tr}

\affiliation{Department of Physics,\\
 Middle East Technical University, 06531 Ankara, Turkey}

\author{Metin Gürses}

\email{gurses@fen.bilkent.edu.tr}

\affiliation{{\small Department of Mathematics, Faculty of Sciences}\\
 {\small Bilkent University, 06800 Ankara, Turkey}}

\author{Tahsin Ça\u{g}r\i{} \c{S}i\c{s}man}

\email{sisman@metu.edu.tr}

\affiliation{Department of Physics,\\
 Middle East Technical University, 06531 Ankara, Turkey}

\author{Bayram Tekin}

\email{btekin@metu.edu.tr}

\affiliation{Department of Physics,\\
 Middle East Technical University, 06531 Ankara, Turkey}

\date{\today}
\begin{abstract}
We give an exact solution of the quadratic gravity in $D$ dimensions.
The solution is a plane-fronted wave metric with a cosmological constant.
This metric solves not only the full quadratic gravity field equations
but also the linearized ones which include the linearized equations
of the recently found critical gravity. A subset of the solutions
change the asymptotic structure of the anti-de Sitter space due to
their logarithmic behavior. 
\end{abstract}


\maketitle
{\small \tableofcontents{}}{\small \par}

\section{Introduction}

Quadratic deformations of Einstein's gravity have always attracted
attention since the inception of general relativity for various reasons.
Initial motivation was to understand the uniqueness of general relativity,
but later on it was realized that perturbative quantum gravity required
these terms \cite{Stelle}. Generically, quadratic gravity theories
at the linearized level describe massless and massive spin-2, and
massive spin-0 modes. Massive spin-2 mode ruins perturbative unitarity
due to its ghost nature. Recently, a new interest in quadratic theories
arose, since it was shown that in three dimensions the ghost disappears
\cite{BHT}, and in $D$-dimensions in the anti-de Sitter (AdS) background,
certain quadratic theories become {}``critical'' with only a massless
spin-2 excitation just like the Einstein gravity \cite{LuPope,DeserLiu}.
These observations led us to consider exact solutions to the quadratic
gravity models. The existence of a cosmological constant changes the
structure of the field equations and the solutions dramatically. For
example, the plane wave metric solves all higher order gravity field
equations coming from string theory with zero cosmological constant
\cite{guv,Deser}. With a cosmological constant this does not work;
however, if one starts with a plane-fronted AdS-wave \cite{gur1,gur2},
one may have exact solutions to quadratic curvature gravity models
as we show below. In general, save the Schwarzschild--anti-de Sitter
solution, no exact solution is known in generic quadratic gravity
theories. In some specific theories, such as the Einstein-Gauss-Bonnet
theory, some static spherically-symmetric solutions are known \cite{boudes};
in the new massive gravity of \cite{BHT}, AdS-wave solutions were
found in \cite{Ayon-Beato}, and types D and N solutions were found
in \cite{AhmedovAliev}. In this paper, in generic $D$-dimensions,
we find the AdS-wave solutions for general quadratic theories including
the critical gravity. By construction, our exact solutions also solve
the linearized wave equations.

The layout of the paper is as follows: In Sec.\ II, we briefly review
the quadratic gravity theory and specifically the critical gravity
in $D$-dimensions. In Sec.\ III, we discuss the AdS-wave metric,
compute the Riemann and the relevant tensors, and derive the field
equations. In Sec.\ IV, we present the solutions of the field equations
deferring one rather cumbersome case to the Appendix.

\section{Critical Points of Quadratic Gravities}

\noindent This work was inspired by the critical gravity models \cite{LuPope,DeserLiu}
which are a certain subset of quadratic curvature gravities. Therefore,
we will briefly recapitulate these models. The action of the quadratic
gravity is \begin{eqnarray}
I & = & \int d^{D}x\,\sqrt{-g}\left[\frac{1}{\kappa}\left(R-2\Lambda_{0}\right)+\alpha R^{2}+\beta R_{\mu\nu}^{^{2}}+\gamma\left(R_{\mu\nu\sigma\rho}^{2}-4R_{\mu\nu}^{2}+R^{2}\right)\right].\label{eq:Quadratic_action}\end{eqnarray}
 The (source-free) field equations were given in \cite{DeserTekin,GulluTekin}
as\begin{align}
\frac{1}{\kappa}\left(R_{\mu\nu}-\frac{1}{2}g_{\mu\nu}R+\Lambda_{0}g_{\mu\nu}\right)+2\alpha R\left(R_{\mu\nu}-\frac{1}{4}g_{\mu\nu}R\right)+\left(2\alpha+\beta\right)\left(g_{\mu\nu}\square-\nabla_{\mu}\nabla_{\nu}\right)R\nonumber \\
+2\gamma\left[RR_{\mu\nu}-2R_{\mu\sigma\nu\rho}R^{\sigma\rho}+R_{\mu\sigma\rho\tau}R_{\nu}^{\phantom{\nu}\sigma\rho\tau}-2R_{\mu\sigma}R_{\nu}^{\phantom{\nu}\sigma}-\frac{1}{4}g_{\mu\nu}\left(R_{\tau\lambda\sigma\rho}^{2}-4R_{\sigma\rho}^{2}+R^{2}\right)\right]\nonumber \\
+\beta\square\left(R_{\mu\nu}-\frac{1}{2}g_{\mu\nu}R\right)+2\beta\left(R_{\mu\sigma\nu\rho}-\frac{1}{4}g_{\mu\nu}R_{\sigma\rho}\right)R^{\sigma\rho} & =0.\label{fieldequations}\end{align}
 The two AdS vacua of the theory satisfy \begin{equation}
\frac{\Lambda-\Lambda_{0}}{2\kappa}+f\Lambda^{2}=0,\qquad f\equiv\left(D\alpha+\beta\right)\frac{\left(D-4\right)}{\left(D-2\right)^{2}}+\gamma\frac{\left(D-3\right)\left(D-4\right)}{\left(D-1\right)\left(D-2\right)}.\label{quadratic}\end{equation}
 Around any of these vacua, the linearized equations read \cite{DeserTekin}
\begin{equation}
c\,\mathcal{G}_{\mu\nu}^{L}+\left(2\alpha+\beta\right)\left(\bar{g}_{\mu\nu}\bar{\square}-\bar{\nabla}_{\mu}\bar{\nabla}_{\nu}+\frac{2\Lambda}{D-2}\bar{g}_{\mu\nu}\right)R^{L}+\beta\left(\bar{\square}\mathcal{G}_{\mu\nu}^{L}-\frac{2\Lambda}{D-1}\bar{g}_{\mu\nu}R^{L}\right)=0,\label{eq:Linearized_eom}\end{equation}
 for the metric perturbation $h_{\mu\nu}\equiv g_{\mu\nu}-\bar{g}_{\mu\nu}$.
Here, $\mathcal{G}_{\mu\nu}^{L}$ is the linearized Einstein tensor,
and $c$ is given by \begin{equation}
c\equiv\frac{1}{\kappa}+\frac{4\Lambda D}{D-2}\alpha+\frac{4\Lambda}{D-1}\beta+\frac{4\Lambda\left(D-3\right)\left(D-4\right)}{\left(D-1\right)\left(D-2\right)}\gamma.\label{eq:c}\end{equation}

The critical theory is obtained as follows: One chooses $4\alpha\left(D-1\right)+D\beta=0$,
which then kills the massive spin-0 mode, and sets $R^{L}=-\frac{2\Lambda}{D-2}h=0$.
Then, in the transverse gauge $\nabla^{\mu}h_{\mu\nu}=0,$ the linearized
equations simplify to\begin{align}
\left(\beta\bar{\square}+c\right)\mathcal{G}_{\mu\nu}^{L} & =0,\label{eq:Critic_short_eqn}\end{align}
or more explicitly \begin{equation}
\left(\bar{\square}-\frac{4\Lambda}{\left(D-1\right)\left(D-2\right)}-M^{2}\right)\left(\bar{\square}-\frac{4\Lambda}{\left(D-1\right)\left(D-2\right)}\right)h_{\mu\nu}=0,\end{equation}
where \begin{equation}
M^{2}=-\frac{1}{\beta}\left(c+\frac{4\Lambda\beta}{\left(D-1\right)\left(D-2\right)}\right),\label{eq:Mass_of_spin-2_excitation}\end{equation}
 and the point $M^{2}=0$ defines the critical point where one is
left only with a massless spin-2 excitation.

\section{$\text{AdS}$-Wave Metric\label{sec:wave-metric}}

The quadratic field equations are highly nontrivial, therefore the
form of the metric ansatz is important in finding solutions. Here,
we take the $D$-dimensional AdS-wave metric (which is conformally
related to the pp-wave metric) to be in the Kerr-Schild form \cite{KerrSchild,GursesGursey}
as\begin{equation}
g_{\mu\nu}=\bar{g}_{\mu\nu}+2V\lambda_{\mu}\lambda_{\nu},\label{denk1}\end{equation}
 where $\bar{g}_{\mu\nu}$ is the metric of the AdS space. The vector
$\lambda^{\mu}=g^{\mu\nu}\lambda_{\nu}=\bar{g}^{\mu\nu}\lambda_{\nu}$
is assumed to be null and geodesic with respect to both $\bar{g}_{\mu\nu}$
and $g_{\mu\nu}$, that is \cite{DereliGurses}\begin{equation}
\lambda_{\mu}\lambda^{\mu}=g^{\mu\nu}\lambda_{\mu}\lambda_{\nu}=\bar{g}^{\mu\nu}\lambda_{\mu}\lambda_{\nu}=0,\qquad\lambda^{\mu}\nabla_{\mu}\lambda^{\nu}=\lambda^{\mu}\bar{\nabla}_{\mu}\lambda^{\nu}=0.\label{eq:null}\end{equation}
 The inverse metric can be found as \begin{equation}
g^{\mu\nu}=\bar{g}^{\mu\nu}-2V\lambda^{\mu}\lambda^{\nu},\end{equation}
 which is an \emph{exact} form. Here, note the similarity with a perturbation
analysis where the metric perturbation is defined as $h_{\mu\nu}\equiv g_{\mu\nu}-\bar{g}_{\mu\nu}$;
and, at the linearized level, the inverse metric becomes $g^{\mu\nu}=\bar{g}^{\mu\nu}-h^{\mu\nu}$.
Now, let us choose the coordinates on AdS to be in the conformally
flat form \begin{equation}
\bar{g}_{\mu\nu}=\phi^{-2}\eta_{\mu\nu},\end{equation}
 where $\phi=k_{\mu}x^{\mu}$, $k_{\mu}$ is a constant vector, and
we choose the flat space coordinates as $x^{\mu}=(u,v,x^{1},\cdots,x^{n})$
with $n=D-2$. This choice of $\bar{g}_{\mu\nu}$ simplifies the construction.
Here, $u$ and $v$ are null coordinates, hence more explicitly\begin{equation}
\eta_{\mu\nu}dx^{\mu}dx^{\nu}=2dudv+\sum_{m=1}^{n}\left(dx^{m}\right)^{2}.\end{equation}
 Then, the vector $k_{\mu}=(0,0,k_{1},\cdots,k_{n})$ is related to
the cosmological constant as \begin{equation}
\bar{R}_{\mu\nu}=\frac{2\Lambda}{D-2}\bar{g}_{\mu\nu},\qquad\Lambda=-\frac{\left(D-1\right)\left(D-2\right)}{2\ell^{2}},\end{equation}
 where $\frac{1}{\ell^{2}}\equiv{\displaystyle \sum_{m=1}^{n}k_{m}k_{m}>0}$,
hence we will be working only with $\Lambda<0$ (note that to conform
with the usual notation, we introduced the AdS radius $\ell$). With
this coordinate choice, $\lambda_{\mu}$ naturally becomes\begin{equation}
\lambda_{\mu}dx^{\mu}=du\Rightarrow\lambda^{\mu}\partial_{\mu}=\phi^{2}\partial_{v}.\end{equation}

The function $V$ is assumed not to depend on $v$, that is \begin{equation}
\lambda^{\mu}\partial_{\mu}V=0.\end{equation}
 This choice is extremely important; since, with it, $h_{\mu\nu}\equiv2V\lambda_{\mu}\lambda_{\nu}$
becomes transverse $\nabla_{\mu}h^{\mu\nu}=\bar{\nabla}_{\mu}h^{\mu\nu}=0$
and traceless $h\equiv g^{\mu\nu}h_{\mu\nu}=\bar{g}^{\mu\nu}h_{\mu\nu}=0$.
Furthermore, as we show below by explicit calculations, the parts
of the curvature tensors which are quadratic in $V$ drop out with
this choice. It is also important to realize that if $V=\frac{c\left(u\right)}{\phi^{2}}$,
then $g_{\mu\nu}$ corresponds to just a coordinate transformed version
of $\bar{g}_{\mu\nu}$. This is clear since in this case, one can
simply define $\left(u,\tilde{v}\right)$ in such a way that the two-dimensional
subspace metric $c\left(u\right)du^{2}+2dudv$ becomes $2dud\tilde{v}$.
This fact will play a role in deciding how our solutions should decay
at infinity.

We are now ready to compute the Riemann tensor. The connection corresponding
to $g_{\mu\nu}$ splits into two parts \begin{equation}
\Gamma_{\alpha\beta}^{\mu}=\bar{\Gamma}_{\alpha\beta}^{\mu}+\Omega_{\phantom{\mu}\alpha\beta}^{\mu},\end{equation}
 where $\bar{\Gamma}_{\alpha\beta}^{\mu}$ is the Levi-Civita connection
corresponding to $\bar{g}_{\mu\nu}$, which reads as \begin{equation}
\bar{\Gamma}_{\alpha\beta}^{\mu}=-\frac{1}{\phi}\left(\delta_{\alpha}^{\mu}\partial_{\beta}\phi+\delta_{\beta}^{\mu}\partial_{\alpha}\phi-\bar{g}_{\alpha\beta}\bar{g}^{\mu\nu}\partial_{\nu}\phi\right).\end{equation}
 The nontrivial part of the connection is given as \begin{equation}
\Omega_{\phantom{\mu}\alpha\beta}^{\mu}=\lambda^{\mu}\left(\lambda_{\alpha}\partial_{\beta}V+\lambda_{\beta}\partial_{\alpha}V+2V\bar{\nabla}_{\beta}\lambda_{\alpha}\right)-\lambda_{\alpha}\lambda_{\beta}\partial^{\mu}V.\end{equation}
 The null vector $\lambda_{\mu}$ with $\partial_{\mu}\lambda_{\nu}=0$
satisfies the following equations, which are frequently used in the
computations:\begin{equation}
\bar{\nabla}_{\nu}\lambda_{\mu}=\frac{1}{\phi}(\lambda_{\mu}\partial_{\nu}\phi+\lambda_{\nu}\partial_{\mu}\phi),\qquad\bar{\square}\lambda_{\mu}=\frac{1-D}{\ell^{2}}\lambda_{\mu},\qquad\bar{g}^{\mu\nu}\left(\bar{\nabla}_{\mu}\lambda_{\alpha}\right)\left(\bar{\nabla}_{\nu}\lambda_{\beta}\right)=\frac{1}{\ell^{2}}\lambda_{\alpha}\lambda_{\beta},\end{equation}
 where $\bar{\square}\equiv\bar{g}^{\mu\nu}\bar{\nabla}_{\mu}\bar{\nabla}_{\nu}$.
The Riemann tensor reduces to\begin{equation}
R_{\phantom{\nu}\alpha\beta\gamma}^{\nu}=\bar{R}_{\phantom{\nu}\alpha\beta\gamma}^{\nu}+\bar{\nabla}_{\beta}\Omega_{\gamma\alpha}^{\nu}-\bar{\nabla}_{\gamma}\Omega_{\beta\alpha}^{\nu}.\end{equation}
 Note that the $\left(\Omega_{\beta\sigma}^{\nu}\Omega_{\gamma\alpha}^{\sigma}-\Omega_{\gamma\sigma}^{\nu}\Omega_{\beta\alpha}^{\sigma}\right)$
part of the Riemann tensor becomes zero, since the explicit calculation
of $\Omega_{\beta\sigma}^{\nu}\Omega_{\gamma\alpha}^{\sigma}$ yields
a symmetric tensor in $\beta$ and $\gamma$ indices\begin{equation}
\Omega_{\beta\sigma}^{\nu}\Omega_{\gamma\alpha}^{\sigma}=-\lambda_{\beta}\lambda_{\gamma}\lambda^{\nu}\lambda_{\alpha}\partial_{\sigma}V\left(\partial^{\sigma}V+\frac{2V}{\phi}\partial^{\sigma}\phi\right).\end{equation}
 After a lengthy computation, one finds the Riemann tensor as\begin{align}
R_{\phantom{\nu}\alpha\beta\gamma}^{\nu}= & \bar{R}_{\phantom{\nu}\alpha\beta\gamma}^{\nu}-\frac{2\lambda^{\nu}}{\phi^{2}}\lambda_{[\beta}\partial_{\gamma]}\partial_{\alpha}\left(V\phi^{2}\right)+\frac{2\lambda^{\nu}}{\phi^{3}}\lambda_{[\beta}\bar{g}_{\gamma]\alpha}\partial_{\rho}\phi\partial^{\rho}\left(V\phi^{2}\right)+\frac{2\lambda_{\alpha}}{\phi^{3}}\lambda_{[\beta}\partial_{\gamma]}\left(V\phi^{2}\right)\partial^{\nu}\phi\nonumber \\
 & +2\lambda_{\alpha}\lambda_{[\beta}\left(\partial_{\gamma]}\partial^{\nu}V+\frac{1}{\phi}\partial_{\gamma]}V\partial^{\nu}\phi-\frac{1}{\phi}\delta_{\gamma]}^{\nu}\partial_{\rho}\phi\partial^{\rho}V\right),\label{eq:Riemann}\end{align}
 where as usual $2A_{[\mu}B_{\nu]}\equiv A_{\mu}B_{\nu}-A_{\nu}B_{\mu}$.
Then, the Ricci tensor $R_{\mu\nu}\equiv R_{\phantom{\rho}\mu\rho\nu}^{\rho}$
and the curvature scalar can be computed as \begin{align}
R_{\mu\nu}= & \frac{2\Lambda}{D-2}g_{\mu\nu}-H\lambda_{\mu}\lambda_{\nu},\qquad R=\frac{2D\Lambda}{D-2},\label{denk2}\end{align}
 where $H$ is defined as \begin{align}
H\equiv & \frac{4}{\phi}\partial_{\rho}V\partial^{\rho}\phi+\bar{\square}V+\frac{4V}{\phi^{2}}\partial_{\rho}\phi\partial^{\rho}\phi+\frac{4\Lambda}{D-2}V.\label{eq:H}\end{align}
 The following two relations will be used in the field equations \begin{align}
\square R_{\mu\nu} & =-\left(\bar{\square}H+\frac{4-2D}{\ell^{2}}H+\frac{4}{\phi}\partial_{\rho}H\partial^{\rho}\phi\right)\lambda_{\mu}\lambda_{\nu},\\
\square\left(V\lambda_{\mu}\lambda_{\nu}\right) & =\left(\bar{\square}V+\frac{4-2D}{\ell^{2}}V+\frac{4}{\phi}\partial_{\rho}V\partial^{\rho}\phi\right)\lambda_{\mu}\lambda_{\nu}.\end{align}

With our metric ansatz, the field Eqs. (\ref{fieldequations}) split
into two parts in the form $Ag_{\mu\nu}+B\lambda_{\mu}\lambda_{\nu}$.
Trace of this equation yields a relation between the effective cosmological
constant and the parameters of the theory exactly given as (\ref{quadratic}),
where we have used $R=\frac{2D\Lambda}{D-2}$ and $R_{\tau\lambda\sigma\rho}^{2}-4R_{\sigma\rho}^{2}+R^{2}=\frac{4\Lambda^{2}D\left(D-3\right)}{\left(D-1\right)\left(D-2\right)}$.
Observe that the $\lambda_{\mu}\lambda_{\nu}$ part does not contribute
to the trace equation. To obtain the rest of the field equations,
the nontrivial computation is the contraction of two Riemann tensors,
that is the $R_{\mu\sigma\rho\tau}R_{\nu}^{\phantom{\nu}\sigma\rho\tau}$
term. After a lengthy computation, one obtains \begin{equation}
R_{\mu\alpha\beta\gamma}R_{\nu}^{\phantom{\nu}\alpha\beta\gamma}=\frac{8\Lambda}{\left(D-1\right)\left(D-2\right)}\left(R_{\mu\nu}-\frac{\Lambda}{D-2}g_{\mu\nu}\right),\end{equation}
 and similarly,\begin{equation}
R_{\mu\sigma}R_{\nu}^{\phantom{\nu}\sigma}=\frac{4\Lambda}{D-2}\left(R_{\mu\nu}-\frac{\Lambda}{D-2}g_{\mu\nu}\right),\qquad R_{\mu\rho\nu\sigma}R^{\rho\sigma}=\frac{2\Lambda}{D-1}\left(R_{\mu\nu}+\frac{2\Lambda}{\left(D-2\right)^{2}}g_{\mu\nu}\right).\end{equation}
 Finally, the remaining field equations become \begin{equation}
\left(\beta\bar{\square}+c\right)\left(H\lambda_{\mu}\lambda_{\nu}\right)=0,\end{equation}
 where $H$ was given in (\ref{eq:H}). Observe the similarity of
this equation to (\ref{eq:Critic_short_eqn}). Then, using\begin{equation}
\bar{\square}\left(\lambda_{\mu}\lambda_{\nu}\right)=\frac{4\Lambda}{D-1}\lambda_{\mu}\lambda_{\nu},\qquad\bar{\nabla}_{\rho}\left(\lambda_{\mu}\lambda_{\nu}\right)=\frac{1}{\phi}\left(2\lambda_{\mu}\lambda_{\nu}\partial_{\rho}\phi+\lambda_{\rho}\lambda_{\nu}\partial_{\mu}\phi+\lambda_{\rho}\lambda_{\mu}\partial_{\nu}\phi\right),\end{equation}
 we get\begin{equation}
\lambda_{\mu}\lambda_{\nu}\beta\left(\bar{\square}+\frac{4}{\phi}\partial^{\sigma}\phi\partial_{\sigma}+\frac{4\left(D-3\right)\Lambda}{\left(D-1\right)\left(D-2\right)}-M^{2}\right)\left(\bar{\square}+\frac{4}{\phi}\partial^{\rho}\phi\partial_{\rho}+\frac{4\left(D-3\right)\Lambda}{\left(D-1\right)\left(D-2\right)}\right)V=0,\label{eq:Field_equations}\end{equation}
 where $M^{2}$ is defined as (\ref{eq:Mass_of_spin-2_excitation}).
Therefore, the exact equations of the quadratic gravity reduces to
a linear fourth-order wave equation. One can show that putting $h_{\mu\nu}=2V\lambda_{\mu}\lambda_{\nu}$
in (\ref{eq:Linearized_eom}) yields (\ref{eq:Field_equations}).
When $M^{2}=0$, the theory reduces to the linearized equations of
the critical theory of \cite{LuPope,DeserLiu}. There is a fine point
here: In the critical theory, to get rid off the massive scalar mode,
one imposes a relation between $\alpha$ and $\beta$, $4\alpha\left(D-1\right)+D\beta=0$;
but here these parameters are arbitrary, since $R_{L}\equiv R-\bar{R}$
vanishes identically for the AdS-wave metric. In the next section,
we discuss the solutions of (\ref{eq:Field_equations}) in detail.

\section{Solutions of the Field Equations\label{sec:Solutions-of-eqns}}

The Eq. (\ref{eq:Field_equations}) is of the form \begin{equation}
\left(\bar{\square}+\frac{4}{\phi}\partial^{\sigma}\phi\partial_{\sigma}+b\right)\left(\bar{\square}+\frac{4}{\phi}\partial^{\rho}\phi\partial_{\rho}+a\right)V=0,\label{eq:Diff_eqns}\end{equation}
 with $a=\frac{4\left(D-3\right)\Lambda}{\left(D-1\right)\left(D-2\right)}$
and $b=a-M^{2}$, which are obviously equal for $M^{2}=0$. Whether
$M^{2}=0$ or not changes the behavior of solutions dramatically,
therefore we will discuss these cases separately.

\subsection{The $a\ne b$ case}

One can define $V=V_{a}+V_{b}$ in such a way that each part satisfies
the corresponding second order equation\begin{equation}
\left(\bar{\square}+\frac{4}{\phi}\partial^{\rho}\phi\partial_{\rho}+a\right)V_{a}=0,\qquad\left(\bar{\square}+\frac{4}{\phi}\partial^{\rho}\phi\partial_{\rho}+b\right)V_{b}=0.\label{eq:Second_order_eqns}\end{equation}
 Without loss of generality, let us choose only the $n^{\text{th}}$
component of the $k^{\mu}$ vector to be nonvanishing; and define
$z=x^{n}$, then $\phi=\frac{z}{\ell}$. With this choice, the quadratic
equation reduces to\begin{equation}
\left(\bar{\square}+\frac{4}{\phi}\partial^{\rho}\phi\partial_{\rho}+a\right)V_{a}=\left(\frac{z^{2}}{\ell^{2}}\partial^{2}+\frac{6-D}{\ell^{2}}z\partial_{z}+a\right)V_{a}=0,\end{equation}
 where $\partial^{2}\equiv\eta^{\mu\nu}\partial_{\mu}\partial_{\nu}$.
Note that the $V_{b}$ equation is similar. Using the separation of
variables technique as $V_{a}\equiv\Psi_{a}\left(u,x^{1},x^{2},\cdots,x^{n-1}\right)\Phi_{a}\left(u,z\right)$,
we can split the quadratic equation into two parts:\begin{align}
\left(\vec{\nabla}^{2}+\chi_{a}\right)\Psi_{a}\left(u,x^{1},x^{2},\cdots,x^{n-1}\right) & =0,\nonumber \\
\left[z^{2}\frac{d^{2}}{dz^{2}}+\left(6-D\right)z\frac{d}{dz}+\left(a\ell^{2}-\chi_{a}z^{2}\right)\right]\Phi_{a}\left(u,z\right) & =0,\label{eq:Seperation_of_vars}\end{align}
 where $\chi_{a}$ is an arbitrary real number at this stage and $\vec{\nabla}^{2}\equiv\sum_{i=1}^{n-1}\frac{\partial^{2}}{\partial\left(x^{i}\right)^{2}}$.
Note that the $\frac{\partial^{2}}{\partial u\partial v}$ term does
not appear because of the $v$-independence of the solution. On the
other hand, the solution will have an arbitrary dependence on $u$.
Depending on the boundary conditions, $\chi_{a}$ can be continuous
or discrete. Then, a \emph{formal} solution will be of the form\begin{align}
V_{a}\left(u,x^{1},x^{2},\cdots,x^{n-1},z\right)= & \int d\chi_{a}\, A(\chi_{a})\Psi_{a}\left(u,x^{1},x^{2},\cdots,x^{n-1};\chi_{a}\right)\Phi_{a}\left(u,z;\chi_{a}\right)\nonumber \\
 & +\sum_{\chi_{a}}{\displaystyle B_{\chi_{a}}\Psi_{a,\chi_{a}}\left(u,x^{1},x^{2},\cdots,x^{n-1}\right)\Phi_{a,\chi_{a}}\left(u,z\right)},\label{eq:Gen_soln}\end{align}
 where $A\left(\chi_{a}\right)$ and $B_{\chi_{a}}$ are arbitrary
functions of $u$. Over the entire $(u,v,x^{i})$ flat space, the
first equation in (\ref{eq:Seperation_of_vars}) does not have bounded
solutions when $\chi_{a}<0$. Therefore, we will take $\chi_{a}\equiv\xi_{a}^{2}\ge0$.
Here, the discussion bifurcates whether $\xi_{a}=0$ or not. Concentrating
on the continuous case, first we start with $\xi_{a}\ne0$. Then,
the solutions are of the form\begin{equation}
\Psi_{a}\left(u,x^{1},x^{2},\cdots,x^{n-1}\right)=c_{1}\left(u\right)\sin(\vec{\xi}_{a}.\vec{r}+c_{2}\left(u\right)),\end{equation}
 where $\vec{\xi}_{a}$ is an arbitrary vector with magnitude $\xi_{a}$,
and $\vec{r}=\left(x^{i}\right)$. Now, we come to the second equation
in (\ref{eq:Seperation_of_vars}) which is in the form of the modified
Bessel equation for $D=5$, and can be converted to this form for
any other $D$ by the following redefinition: \begin{equation}
\Phi_{a}\left(u,z\right)\equiv z^{\frac{D-5}{2}}f_{a}\left(u,z\right),\end{equation}
 which then yields\begin{equation}
\left[z^{2}\frac{d^{2}}{dz^{2}}+z\frac{d}{dz}-\left(\nu_{a}^{2}+\xi_{a}^{2}z^{2}\right)\right]f_{a}\left(u,z\right)=0,\end{equation}
 where $\nu_{a}=\frac{1}{2}\sqrt{\left(D-5\right)^{2}-4a\ell^{2}}$.
For generic $D$ and nonvanishing $\xi_{a}$, the solution is given
as \begin{equation}
\Phi_{a}\left(u,z\right)=z^{\frac{D-5}{2}}\left[c_{3}\left(u\right)I_{\nu_{a}}\left(z\xi_{a}\right)+c_{4}\left(u\right)K_{\nu_{a}}\left(z\xi_{a}\right)\right],\end{equation}
 where $I_{\nu}$ and $K_{\nu}$ are the modified Bessel functions
of the first and second kind, respectively. $V_{b}$ will have the
similar solutions with $\nu_{b}=\frac{1}{2}\sqrt{\left(D-5\right)^{2}-4b\ell^{2}}=\frac{1}{2}\sqrt{\left(D-1\right)^{2}+4\ell^{2}M^{2}}$.
As we discussed in Sec. \ref{sec:wave-metric}, we require the solution
to go like $\frac{1}{z^{2}}$ at the boundary $z=0$; the modified
Bessel functions approach $z\rightarrow0$ as $I_{\nu}\left(z\right)\sim z^{\nu}$,
$K_{\nu}\sim z^{-\nu}$ and $K_{0}\sim-\ln z$. Therefore, we keep
both $c_{3}$ and $c_{4}$. It is important to realize that $\nu_{a}$
and $\nu_{b}$ are real. This is automatically satisfied for $\nu_{a}$
since $a=-\frac{2\left(D-3\right)}{\ell^{2}}$, and $\nu_{a}=\pm\frac{1}{2}\left(D-1\right)$.
The reality of $\nu_{b}$ puts a constraint on $M^{2}$ which is\begin{equation}
M^{2}\ge-\frac{\left(D-1\right)^{2}}{4\ell^{2}}.\label{eq:bf}\end{equation}
 This bound is exactly equivalent to the Breitenlohner-Freedman (BF)
bound on the mass of a scalar excitation in AdS \cite{BF}. When the
bound is saturated, $\nu_{b}=0$ and logarithmic solutions arise.

For the sake of completeness, let us write the solution\begin{align}
V\left(u,\vec{r},z\right)= & z^{\frac{D-5}{2}}\left[c_{a,1}\left(u\right)I_{\nu_{a}}\left(z\xi_{a}\right)+c_{a,2}\left(u\right)K_{\nu_{a}}\left(z\xi_{a}\right)\right]\sin(\vec{\xi_{a}}.\vec{r}+c_{a,3}\left(u\right))\nonumber \\
 & +z^{\frac{D-5}{2}}\left[c_{b,1}\left(u\right)I_{\nu_{b}}\left(z\xi_{b}\right)+c_{b,2}\left(u\right)K_{\nu_{b}}\left(z\xi_{b}\right)\right]\sin(\vec{\xi_{b}}.\vec{r}+c_{b,3}\left(u\right)).\label{eq:Gen_soln_ksi-ne-0}\end{align}
The full solution can be obtained by integrating the first part of
(\ref{eq:Gen_soln_ksi-ne-0}) with respect to $\vec{\xi}_{a}$, and
the second part with respect to $\vec{\xi}_{b}$ as in (\ref{eq:Gen_soln}).
Here, $c_{a,1}\left(u\right)$ and $c_{a,2}\left(u\right)$ depend
on $\vec{\xi}_{a}$, and $c_{b,1}\left(u\right)$ and $c_{b,2}\left(u\right)$
depend on $\vec{\xi}_{b}$. In odd dimensions for $D\ge5$, $c_{2,a}$
should vanish, otherwise $V$ is unbounded at the boundary of AdS
at $z=0$. (Note that we do not worry about the disconnected component
of the boundary, which is just the single point $z=\infty$, since
this point just compactifies the boundary of AdS from $\mathbb{R}^{D-1}\rightarrow S^{D-1}$.)
For generic values of $M^{2}$, $\nu_{b}$ is not an integer or an
odd integer, therefore in general $c_{2,b}$ must be kept.

Let us now consider the $\xi_{a}=\xi_{b}=0$ case. {[}The case when
only one of these parameters vanish follows from the discussion above
and the discussion below.{]} In this case, $\Psi_{a}\left(u,x^{1},x^{2},\cdots,x^{n-1}\right)=c\left(u\right)+\vec{q}\left(u\right).\vec{r}$,
but we take $\vec{q}\left(u\right)=0$ to have a bounded solution
at $\left|\vec{r}\right|\rightarrow\infty$. The other equation reduces
to \begin{equation}
\left(z^{2}\frac{d^{2}}{dz^{2}}+z\frac{d}{dz}-\nu_{a}^{2}\right)f_{a}\left(u,z\right)=0,\end{equation}
 whose solution is $f_{a}\left(z\right)=c_{a,1}\left(u\right)z^{\left|\nu_{a}\right|}+c_{a,1}\left(u\right)z^{-\left|\nu_{a}\right|}$.
Adding also the solution of the $b$ equation, one gets\begin{equation}
V\left(u,z\right)=c_{a,1}\left(u\right)z^{D-3}+c_{a,2}\left(u\right)\frac{1}{z^{2}}+z^{\frac{D-5}{2}}\left(c_{b,1}\left(u\right)z^{\left|\nu_{b}\right|}+c_{b,2}\left(u\right)z^{-\left|\nu_{b}\right|}\right).\label{eq:ksi-zero}\end{equation}
When $M^{2}>0$, because of the last term the spacetime is not asymptotically
AdS as one approaches the boundary $z=0$: Namely, $V\left(u,z\right)\sim c_{b,2}\left(u\right)z^{-\left(2+\epsilon\right)}$
where $\epsilon>0$, hence $c_{b,2}\left(u\right)=0$. On the other
hand, when $0>M^{2}>-\frac{\left(D-1\right)^{2}}{4\ell^{2}}$, all
the terms are allowed. When the BF bound (\ref{eq:bf}) is saturated,
one has ($\nu_{b}=0$) \begin{equation}
V\left(u,z\right)=c_{a,1}\left(u\right)z^{D-3}+c_{a,2}\left(u\right)\frac{1}{z^{2}}+z^{\frac{D-5}{2}}\left[c_{b,1}\left(u\right)+c_{b,2}\left(u\right)\ln\left(\frac{z}{\ell}\right)\right],\label{eq:ksi-zero_BF}\end{equation}
 which yields an asymptotically AdS metric. 

Up to now, we have implicitly assumed $D>3$, but in fact our expressions
are also valid for $D=3$ for which (\ref{eq:ksi-zero}) and (\ref{eq:ksi-zero_BF})
reduce to the results given in \cite{Ayon-Beato}.

\subsection{The $a=b$ case which includes the critical theory}

In this case, $a=-\frac{2\left(D-3\right)}{\ell^{2}}$ and $V_{a}$
, as found above (\ref{eq:Gen_soln_ksi-ne-0}), is \emph{a} solution,
but this is not the only solution: One should consider the full quadratic
theory; \begin{equation}
\left(z^{2}\partial^{2}+\left(6-D\right)z\partial_{z}-2\left(D-3\right)\right)^{2}V=0.\end{equation}
 Defining $W\equiv\left(z^{2}\partial^{2}+\left(6-D\right)z\partial_{z}-2\left(D-3\right)\right)V$;
so that, $\left(z^{2}\partial^{2}+\left(6-D\right)z\partial_{z}-2\left(D-3\right)\right)W=0$,
since we know the solution of the latter equation from the above discussion,
we can simply consider the nonhomogeneous equation, where $W$ is
a source term. The $\xi\ne0$ case is somewhat cumbersome and not
particularly illuminating, therefore we defer it to the Appendix,
and here study the $\xi=0$ case. Then, the solution to the quadratic
equation is \begin{equation}
W\left(u,z\right)=c_{1}\left(u\right)z^{D-3}+c_{2}\left(u\right)z^{-2}.\end{equation}
 The nonhomogeneous equation becomes\begin{equation}
\left(z^{2}\partial^{2}+\left(6-D\right)z\partial_{z}-2\left(D-3\right)\right)V=c_{1}\left(u\right)z^{D-3}+c_{2}\left(u\right)z^{-2},\end{equation}
 which after rescaling $V\left(u,z\right)=z^{\frac{D-5}{2}}f\left(u,z\right)$
can be transformed to \begin{equation}
\left(z^{2}\frac{d^{2}}{dz^{2}}+z\frac{d}{dz}-\frac{\left(D-1\right)^{2}}{4}\right)f\left(z\right)=c_{1}\left(u\right)z^{\frac{D-1}{2}}+c_{2}\left(u\right)z^{\frac{1-D}{2}}.\end{equation}
 The general solution of this equation is\begin{equation}
V\left(u,z\right)=d_{1}\left(u\right)z^{D-3}+\frac{d_{2}\left(u\right)}{z^{2}}+\frac{1}{D-1}\left(c_{1}\left(u\right)z^{D-3}-\frac{c_{2}\left(u\right)}{z^{2}}\right)\ln\left(\frac{z}{\ell}\right),\end{equation}
 which was also obtained recently in \cite{Alishah}, in the context
of $D$-dimensional Log gravity. For generic $D$, the solution is
not asymptotically AdS, unless $c_{2}\left(u\right)$ vanishes.\textcolor{red}{{}
}For $D=3$, this equation again reduces to the corresponding expression
given in \cite{Ayon-Beato}.

\section{Conclusions}

We have found exact AdS-wave solutions in the generic quadratic gravity
theory with a cosmological constant. The metrics we have found also
solve the linearized field equations of the same theory. When we restrict
the quadratic theory by choosing $M^{2}=0$, which boils down to eliminating
one of the parameters of the quadratic theory, the solutions we found
in this case also solve the critical gravity theory defined recently.
Depending on the value of $M^{2}$, asymptotic behavior of the solution
changes dramatically. Energy and some other physical properties of
our solutions, and their conformal field theory duals need to be investigated.
It would also be interesting to take the solutions presented here
as background and study the spin-2 fluctuations. 

Finally, with adjusted $\Lambda$ and $M^{2}$, the metrics we have
found will also solve any higher (cubic or more) curvature gravity
models and constitute an example of spacetimes studied in \cite{Hervik}.
This can be seen as follows: the linearized version of a generic gravity
theory constructed from the contractions of the Riemann tensor around
AdS will be exactly of the form (\ref{eq:Linearized_eom}) which is
solved by the AdS-wave metrics obtained above. Since by construction
the exact field equations reduce to the linearized equations for these
solutions, AdS-wave solves the full nonlinear theory at any order
in the curvature.

\section{\label{ackno} Acknowledgments}

M. G. is partially supported by the Scientific and Technological Research
Council of Turkey (TÜB\.{I}TAK) and Turkish Academy of Sciences (TÜBA).
The work of I.G., T.C.S., B.T. is supported by the TÜB\.{I}TAK Grant
No.\ 110T339, and METU Grant No.\ BAP-07-02-2010-00-02.

\appendix

\section{Solution for the Critical Theory with $\xi\ne0$}

Let us consider the $a=b$ theory for the $\xi\ne0$ case. The corresponding
fourth-order equation reduces to the quadratic nonhomogeneous equation;\begin{align}
\left(z^{2}\partial^{2}+\left(6-D\right)z\partial_{z}-2\left(D-3\right)\right)V= & z^{\frac{D-5}{2}}\left[c_{1}\left(u\right)I_{\nu}\left(z\xi\right)+c_{2}\left(u\right)K_{\nu}\left(z\xi\right)\right]\label{eq:Diff_eqn-a_equal_b}\\
 & \times\left[c_{3}\left(u\right)e^{i\vec{\xi}.\vec{r}}+c_{4}\left(u\right)e^{-i\vec{\xi}.\vec{r}}\right],\nonumber \end{align}
 where $\nu=\pm\frac{1}{2}\left(D-1\right)$, but we shall restrict
to the positive $\nu$ case; and instead of the sines and cosines
we choose the exponentials. This equation can be solved with the help
of the Green's function technique. First, we would like to take care
of the $\vec{r}$ dependence using the Fourier transform (for the
sake of simplicity, here we choose $\xi$ to be continuous, but the
discrete case follows similarly) \begin{equation}
V\left(u,z,\vec{r}\right)=\frac{1}{\left(2\pi\right)^{\frac{D-3}{2}}}\int d^{D-3}p\,\tilde{V}\left(u,z,\vec{p}\right)e^{i\vec{p}.\vec{r}},\end{equation}
 then (\ref{eq:Diff_eqn-a_equal_b}) after defining $\tilde{V}\left(u,z,\vec{p}\right)=z^{\frac{D-5}{2}}f\left(u,z,\vec{p}\right)$,
reduces to\begin{align}
\left[z^{2}\frac{d^{2}}{dz^{2}}+z\frac{d}{dz}-\left(\nu^{2}+p^{2}z^{2}\right)\right]f\left(u,z,\vec{p}\right)= & \left(2\pi\right)^{\frac{D-3}{2}}\left[c_{1}\left(u\right)I_{\nu}\left(z\xi\right)+c_{2}\left(u\right)K_{\nu}\left(z\xi\right)\right]\label{eq:z-eqn}\\
 & \times\left[c_{3}\left(u\right)\delta\left(\vec{\xi}-\vec{p}\right)+c_{4}\left(u\right)\delta\left(\vec{\xi}+\vec{p}\right)\right].\nonumber \end{align}
 From the solutions of the homogeneous part, we can construct the
Green's function $\left(\mathcal{O}G=-1\right)$ as\begin{equation}
G\left(z,z^{\prime};p\right)=\frac{1}{z^{\prime}}\begin{cases}
I_{\nu}\left(zp\right)K_{\nu}\left(z^{\prime}p\right) & 0<z<z^{\prime},\\
I_{\nu}\left(z^{\prime}p\right)K_{\nu}\left(zp\right) & z^{\prime}<z<\infty,\end{cases}\end{equation}
 where we have used the Wronskian $W\left\{ I_{\nu}\left(pz\right)K_{\nu}\left(pz\right)\right\} =-\frac{1}{z}$.
Therefore, the solution of (\ref{eq:z-eqn}) becomes\begin{align}
f\left(u,z,\vec{p}\right)= & \left[d_{1}\left(u\right)I_{\nu}\left(zp\right)+d_{2}\left(u\right)K_{\nu}\left(zp\right)\right]\nonumber \\
 & +\left(2\pi\right)^{\frac{D-3}{2}}\left[c_{3}\left(u\right)\delta\left(\vec{\xi}-\vec{p}\right)+c_{4}\left(u\right)\delta\left(\vec{\xi}+\vec{p}\right)\right]\nonumber \\
 & \phantom{+}\times\int_{0}^{\infty}dz^{\prime}\, G\left(z,z^{\prime};p\right)\left[c_{1}\left(u\right)I_{\nu}\left(z^{\prime}\xi\right)+c_{2}\left(u\right)K_{\nu}\left(z^{\prime}\xi\right)\right].\end{align}
 We can carry out the $p$ integrals using \begin{equation}
\int d^{D-3}p\, f\left(p\right)e^{i\vec{p}.\vec{r}}=\left(2\pi\right)^{\frac{D-3}{2}}r^{\frac{5-D}{2}}\int_{0}^{\infty}dp\: f\left(p\right)p^{\frac{D-3}{2}}J_{\left(D-5\right)/2}\left(pr\right),\end{equation}
 where $J_{n}$ is the Bessel function of the first kind. Since the
Fourier transform of $I_{\nu}\left(zp\right)$ diverges, we must choose
$d_{1}\left(u\right)=0$. For $D=4$ and $D=5$, $d_{2}\left(u\right)$
must also be zero, since the integral involving $K_{\nu}\left(zp\right)$
diverges. Then, the integral involving $K_{\nu}\left(zp\right)$ for
$D>5$ gives $z^{-\nu}\,_{2}\tilde{F}_{1}\left(\frac{\nu}{2}-1,\frac{3\nu}{2}-1,\nu-1,-\frac{r^{2}}{z^{2}}\right)$
where the second factor is the regularized hypergeometric function.
As $z\rightarrow0$, this expression diverges. Therefore, for all
$D$, there is no contribution from the homogeneous part, and $c_{2}\left(u\right)$
should vanish in the nonhomogeneous part since that term diverges
in the $\left[0,z\right]$ integral, yielding finally \begin{align}
V\left(u,z,\vec{r}\right)= & z^{\frac{D-5}{2}}\left[c_{1}\left(u\right)e^{i\vec{\xi}.\vec{r}}+c_{2}\left(u\right)e^{-i\vec{\xi}.\vec{r}}\right]\int_{0}^{\infty}dz^{\prime}\, G\left(z,z^{\prime};\xi\right)I_{\nu}\left(z^{\prime}\xi\right)\nonumber \\
= & z^{\frac{D-5}{2}}\left[c_{1}\left(u\right)e^{i\vec{\xi}.\vec{r}}+c_{2}\left(u\right)e^{-i\vec{\xi}.\vec{r}}\right]\\
 & \times\left[K_{\nu}\left(z\xi\right)\int_{0}^{z}\frac{1}{z^{\prime}}I_{\nu}\left(z^{\prime}\xi\right)I_{\nu}\left(z^{\prime}\xi\right)\, dz^{\prime}+I_{\nu}\left(z\xi\right)\int_{z}^{\infty}\frac{1}{z^{\prime}}K_{\nu}\left(z^{\prime}\xi\right)I_{\nu}\left(z^{\prime}\xi\right)\, dz^{\prime}\right].\nonumber \end{align}
This solution is valid for a given $\vec{\xi}$. The general solution
can be obtained by integrating this solution over the $\left(D-3\right)$-dimensional
$\vec{\xi}$-space. Here, $c_{1}\left(u\right)$ and $c_{2}\left(u\right)$
also depend on $\vec{\xi}$. Using Mathematica, one can find the integrals
in terms of the hypergeometric function $\,_{p}F_{q}\left(a;b;z\right)$
and the digamma function $\psi$ \begin{align}
\int_{0}^{z}dz^{\prime}\,\frac{1}{z^{\prime}}I_{\nu}\left(z^{\prime}\xi\right)I_{\nu}\left(z^{\prime}\xi\right)= & \frac{\left(\xi z\right)^{2\nu}}{2^{2\nu+1}\nu\left[\Gamma\left(\nu+1\right)\right]^{2}}\,_{2}F_{3}\left(\nu,\nu+\frac{1}{2};\nu+1,\nu+1,2\nu+1;\xi^{2}z^{2}\right),\\
\int_{z}^{\infty}\frac{1}{z^{\prime}}K_{\nu}\left(z^{\prime}\xi\right)I_{\nu}\left(z^{\prime}\xi\right)\, dz^{\prime}= & \frac{\xi^{2}z^{2}}{8\nu\left(\nu^{2}-1\right)}\,_{3}F_{4}\left(1,1,\frac{3}{2};2,2,2-\nu,\nu+2;\xi^{2}z^{2}\right)\nonumber \\
 & -\frac{\Gamma\left(-\nu\right)\left(\xi z\right)^{2\nu}}{4^{\nu+1}\nu\Gamma\left(\nu+1\right)}\,_{2}F_{3}\left(\nu,\nu+\frac{1}{2};\nu+1,\nu+1,2\nu+1;\xi^{2}z^{2}\right)\nonumber \\
 & -\frac{1}{2\nu}\left[\ln\left(\frac{z\xi}{2}\right)-\psi\left(\nu\right)-\frac{1}{2\nu}\right].\end{align}
 Specifically, for $D=4$, that is $\nu=\frac{3}{2}$, around $z=0$,
one has \begin{equation}
V\left(u,z,\vec{r}\right)\sim\left[c_{1}\left(u\right)e^{i\vec{\xi}.\vec{r}}+c_{2}\left(u\right)e^{-i\vec{\xi}.\vec{r}}\right]\xi^{3/2}z\left[\ln\left(\xi z\right)-1.3963\right],\end{equation}
 which gives an asymptotically AdS metric. 

\newpage{}


\begin{thebibliography}{19}
\bibitem{Stelle} K.~S.~Stelle, Phys. Rev. \textbf{D16}, 953 (1977);
Gen. Rel. Grav. \textbf{9}, 353 (1978).

\bibitem{BHT} E.~A.~Bergshoeff, O.~Hohm and P.~K.~Townsend,
Phys.\ Rev.\ D\textbf{ 79}, 124042 (2009).

\bibitem{LuPope} H.~Lu and C.~N.~Pope, {}``Critical Gravity in
Four Dimensions,'' arXiv:1101.1971 {[}hep-th{]}.

\bibitem{DeserLiu} S.~Deser, H.~Liu, H.~Lu, C.~N.~Pope, T.~C.~Sisman
and B.~Tekin, Phys.\ Rev.\ D \textbf{83}, 061502 (2011) .

\bibitem{guv} R.~G{ü}ven, Phys.\ Lett.\ B \textbf{191}, 275
(1987).

\bibitem{Deser} S.~Deser, J.\ Phys.\ A \textbf{8}, 1972 (1975). 

\bibitem{gur1} M.~G{ü}rses, J.\ Phys.\ A \textbf{14}, 1957 (1981).

\bibitem{gur2} P.~Baekler and M.~G{ü}rses, Lett.\ Math.\ Phys.\ \textbf{14},
185 (1987).

\bibitem{boudes} D.~G.~Boulware and S.~Deser, Phys.\ Rev.\ Lett.\ \textbf{55},
2656 (1985).

\bibitem{Ayon-Beato} E.~Ayon-Beato, G.~Giribet and M.~Hassaine,
JHEP \textbf{0905}, 029 (2009).

\bibitem{AhmedovAliev} H.~Ahmedov and A.~N.~Aliev, Phys.\ Lett.\ B
\textbf{694}, 143 (2010); Phys.\ Rev.\ Lett.\ \textbf{106}, 021301
(2011).

\bibitem{DeserTekin} S.~Deser and B.~Tekin, Phys.\ Rev.\ Lett.\textbf{\ 89},
101101 (2002); Phys.\ Rev.\ D \textbf{67}, 084009 (2003).

\bibitem{GulluTekin} I.~Gullu and B.~Tekin, Phys.\ Rev.\ D \textbf{80},
064033 (2009).

\bibitem{KerrSchild} R.~P.~Kerr and A.~Schild, Proc.\ Symp.\ Appl.\ Math.\textbf{\ 17
}199\textbf{ }(1965); G.~C.~Debney, R.~P.~Kerr and A.~Schild
, J.\ Math.\ Phys.\textbf{\ 10} 1842 (1969). 

\bibitem{GursesGursey} M.~Gürses and F.~Gürsey, J.\ Math.\ Phys.\ \textbf{16}
2385 (1975).

\bibitem{DereliGurses} T.~Dereli and M.~G{ü}rses, Phys.\ Lett.\ \textbf{B171},
209-211 (1986).

\bibitem{BF} P.~Breitenlohner and D.~Z.~Freedman, Phys.\ Lett.\ B
\textbf{115}, 197 (1982).

\bibitem{Alishah} M.~Alishahiha and R.~Fareghbal, {}``D-Dimensional
Log Gravity,'' arXiv:1101.5891 {[}hep-th{]}. 

\bibitem{Hervik} A.~A.~Coley, G.~W.~Gibbons, S.~Hervik and C.~N.~Pope,
Class.\ Quant.\ Grav.\ \textbf{25}, 145017 (2008).
\end{thebibliography}
\end{document}